# Construction of Non-Perturbative, Unitary Particle-Antiparticle Amplitudes for Finite Particle Number Scattering Formalisms[#]

James Lindesay[*,+] and H. Pierre Noyes[*]

[*]Stanford Linear Accelerator Center, Stanford University, Stanford, California 94309
[+]Computational Physics Laboratory, Howard University, Washington, D.C. 20059

**Abstract**

Starting from a unitary, Lorentz invariant two-particle scattering amplitude , we show how to use an identification and replacement process to construct a unique, unitary particle-antiparticle amplitude. This process differs from conventional on-shell Mandelstam s,t,u crossing in that the input and constructed amplitudes can be off-diagonal and off-energy shell. Further, amplitudes are constructed using the invariant parameters which are appropriate to use as driving terms in the multi-particle, multichannel non-perturbative, cluster decomposable, relativistic scattering equations of the Faddeev-type integral equations recently presented by Alfred, Kwizera, Lindesay and Noyes. It is therefore anticipated that when so employed, the resulting multi-channel solutions will also be unitary. The process preserves the usual particle-antiparticle symmetries. To illustrate this process, we construct a J=0 scattering length model chosen for simplicity. We also exhibit a class of physical models which contain a finite quantum mass parameter and are Lorentz invariant. These are constructed to reduce in the appropriate limits, and with the proper choice of value and sign of the interaction parameter, to the asymptotic solution of the non-relativistic Coulomb problem, including the forward scattering singularity , the essential singularity in the phase, and the Bohr bound-state spectrum.

PACS: 11.80.-m, 11.80.Cr, 11.80.Jy

## I. INTRODUCTION

This paper is part of a research program aimed at constructing a general, unitary, *non-perturbative* *N*-particle relativistic scattering theory. The *N*- particle amplitude must always be decomposable into a sum over all possible decompositions into a spectator cluster containing *n* particles which only enter this part of the problem kinematically and a dynamical cluster within which the *m=N-n* remaining particles are described as a fully interacting (quantum entangled) *m*-particle system. The input driving the relativistic

---

[#] Work supported in part by Department of Energy contract DE-AC03-76SF00515

Faddeev-type integral equations are the *m*-particle unitary amplitudes describing all possible *m*-particle directly interacting sub-systems. The basic fact that this theory can be constructed, explicitly formulated, and shown to yield calculable amplitudes that are both unitary and Lorentz invariant has already been proven [1, 2, 3, 4, 5]. It remains to demonstrate that this *fixed* particle number formulation of relativistic scattering theory can be extended to include anti-particles and quanta.

Scattering theory as derived from Hamiltonians has been used as a powerful tool for describing a variety of physical processes. The formalism describes the eigenstates of a fully interacting system of particles with a well defined energy in terms of eigenstates of solvable systems which have an overlapping spectrum of eigenvalues. For our purposes, the most convenient eigenstates to use are the boundary states, which satisfy the incoming or outgoing state asymptotic form, but are otherwise only self-interacting. This means that their masses, charges, and other parameters will have physical values, and as such will not require "renormalization" or "dressing." One is able to appropriately extract bound state systems in the final state directly out of scattering amplitudes from which the cross sections for physical processes can be calculated, as well as explore unitarity and transformation behavior analytically in regimes for which perturbative methods would not be applicable. Since the amplitudes are described in terms of the actual boundary states, only a finite number of degrees of freedom need to be considered to calculate a given scattering process.

The introduction of antiparticles into this form of scattering theory has been problematic in several ways. First, since particle and antiparticle pairs can annihilate or be created, particle number is no longer conserved in a naive way, and particle *nature* can change in a way that is not naively consistent with scattering theoretic approaches. Also, there is no obvious non-relativistic analogue of a transformation or an annihilation process, which are inherently a relativistic. Typically, one's intuition in scattering theoretic approaches is driven by the non-relativistic scattering theory which works so well in describing many low energy processes, but always involves a persistence of any constituent particles, even if their clusterings do change. Considerable effort [6, 7] has been put into guaranteeing unitary outcomes from S matrix approaches to the description of scattering processes. Unitarity is one of the most useful properties of these approaches, and we do not give up on satisfying this property.

The form of the scattering formalism that the authors have found to be most fruitful in this pursuit adapts the non-relativistic framework developed by Faddeev [8] into a relativistically invariant form [1-5]. In this formalism, the scattering amplitude is decomposed into various clusters which are summed over; if they are properly embedded, the unitarity of the total scattering amplitude is guaranteed from the unitarity of the input amplitudes. Using this formalism, one is able to properly traverse relativistic thresholds and demonstrably maintain unitarity when examining rearrangement scattering and breakup of relativistic clusters [2]. This incorporation of production thresholds succeeds because of the multi-channel nature of the few-particle scattering formalism. Antiparticle identifications have yet to be incorporated into such an approach. The type of two-particle present here an approach that can later be incorporated into a few-particle formalism which then will include antiparticles and describe pair creation in a way that allows particle number to change, but remain finite.

We show that, given an exactly unitary and Lorentz invariant particle-particle amplitude, we can always unambiguously construct the corresponding unitary and Lorentz-invariant particle-antiparticle amplitude with the expected particle-antiparticle symmetries. Our approach achieves cluster decomposability by requiring that the dynamical cluster and the full system *both* conserve their individual Lorentz coordinate frames ; this helps insure that the spectating cluster coordinates enter the problem kinematically rather than dynamically. This in turn requires us to use 4-velocity rather than 4-momentum transformations between Faddeev channels. The result is that the natural variables for us are *not* the familiar Mandelstam invariants *(s,t,u)* but instead the invariant energy, angle variables used in [1].
In section II take care to showing the relationship between the two descriptions and in
exhibiting what form ``*st, su,* and *tu* crossing" take in our variables. It is particularly important to realize that in our formalism we must carefully distinguish between amplitudes which are *off-diagonal*, in the sense that they connect 4-momenta in the initial and final states in a way that does not occur for the values of these variables that describe the physically observable processes, and amplitudes which are *off-shell* (i.e. refer to energy values that cannot be reached from the initial boundary state).

Section III then can construct particle-antiparticle amplitudes in two ways starting from a particle-particle amplitude (pp channel) by means of an explicit identification of which particle changes to an anti-particle. The first way, which is closely related to *s-u* crossing, defines the particle-antiparticle scattering amplitude (p-pbar channel). The second way, related to *s-t* crossing, defines the transformation amplitude (X) channel and has no non-relativistic analog. Both replacements preserve the appropriate symmetries. We find that, when expressed in terms of our choice of Lorentz invariant parameters, the three possible amplitudes exhibit *form invariance*. This greatly simplifies subsequent discussion. In particular, this allows us to start from any one amplitude as given and construct the other two, allowing us to drop the requirement of starting from the particle-particle amplitude when it comes to creating models.

In the fourth section we prove that the appropriate amplitudes so constructed are, indeed, unitary. This involves some interesting subtleties when it comes to showing why and how the transformation (X) channel is connected to the particle-antiparticle scattering channel, and how this in turn relates to the problems posed by the coherence of these amplitudes for identical particles.

Our fifth section presents two explicit models that illustrate how the replacement works out and indeed do produce unitary amplitudes for the two-body particle-antiparticle problem. The first is simply an s-wave scattering length model, which has the virtue of producing results that can be easily checked without getting bogged down in the formalism. The second is more physical in that it starts from the solution of the non-relativistic Coulomb problem and produces relativistic generalizations using a finite mass quantum exchange. In the zero quantum mass limit, all these models preserve the forward scattering singularity characteristic of Rutherford scattering and the essential singularity of the Coulomb phases. When the interaction parameter has the appropriate sign and value, they also yield the non-relativistic Bohr bound-state spectrum in the appropriate limits. When we replace our model particle-particle amplitude by a transformation (X) amplitude, we find that this contains an ``exchanged quantum'' which indeed has a deep-lying singularity (compared to elastic scattering threshold) at a relativistic energy corresponding to the quantum mass; this singularity also goes to zero mass in the same limit that gives Coulomb scattering in both the particle-particle and particle-antiparticle channels.

The fact that the quantum singularity in the transformation (X) channel shares so many characteristics with a bound state, and that neither can be reached from a two body input channel without first embedding our model in a three or more particle space, might lead us to call it a ``particle-antiparticle bound state''. But this would be incorrect if taken too literally. Embedded in a larger space, the fact that this quantum carries kinematic variables but need carry no conserved quantum numbers allows it to couple to *any* particle-antiparticle pair allowed in the larger space. In contrast, a *composite* bound state can only be taken apart into its constituents. Thus in our ``finite particle number theory'' the distinction between ``particles'' and ``quanta'' may turn out to be well defined. Nevertheless, as we discuss in the concluding section, we expect that when our two body model for a particle-antiparticle amplitude is embedded in a three particle space we will be able to use our formalism to treat either a finite or a zero mass quantum as a boundary state in that space. The implied possibility of using our theory to describe quantum-particle scattering, pair creation and quantum emission and absorption in a unitary way will be explored elsewhere. It is the fact that our formalism and construction is throughout informed by the necessity (for us) of keeping this possibility alive that leads to some of the complexity of approach. It is this goal that allows us to ask the reader to be patient with what at first sight might appear to be a cumbersome formalism.

## II. DESCRIPTION OF VARIABLES AND PARAMETERS

We begin this section by describing the relevant 2-particle parameters we use in the description of the scattering amplitudes. For present purposes, scattering processes are assumed to involve scalar (spinless) particles. The most general form for the scattering transition amplitude T satisfies integral equation relations in several variables, representing the interacting system in terms of the kinematic parameters (for instance, invariant energy and orientation angles) corresponding to a complete set of boundary states (which are defined to satisfy the boundary conditions of the asymptotic form of the system. Typically, the integral equations satisfied by the amplitudes involve terms which are off-diagonal in the kinematics of the boundary state expansion (M'≠M), as well as generally being off-shell in the (sometimes complex) system energy parameter Z, where M' and M are Lorentz-invariant center-of-momentum energy parameters for the boundary states, and Z is the same parameter for the overall system. Describing the invariant direction unit vectors of the internal pair momenta in the center-of-momentum system by $\hat{q}'$ and $\hat{q}$, this general off-diagonal, off-shell transition amplitude is symbolized by

$$T(M',\hat{q}'|M,\hat{q};Z).$$

If one examines the analytic behavior of this amplitude in the complex Z plane, there will be distinct poles in the Z dependence of the amplitude corresponding to any bound states which the system might have, as well as a forward scattering cut guaranteeing a discontinuity which distinguishes incoming states from outgoing states (necessary for the unitarity prescription). The scattering cut results from the region of kinematic overlap in the eigenvalue spectra of the boundary states and the fully dynamical interacting system. In what follows, we will be careful to distinguish between the off-diagonal behavior of the matrix elements of the scattering amplitudes, the off-shell behavior with regards to the eigenvalue parameter Z, and the analytic behavior of the fully on-shell amplitude $T(M, \hat{q}'| M, \hat{q}; M)$ in terms of the parameter M, which clearly will mix up these analytic behaviors.

Our goal will be to describe Lorentz invariant amplitudes for various physical scattering processes which have relationships between them that will guarantee unitarity. We will attempt to describe scattering processes between particles that have conserved particle quantum numbers, and as such cannot be singly produced in any physical process. Quanta as such would only mediate the interaction between these particles, although generally one can describe interactions without the necessity of introducing quanta. The problem then becomes one of defining the interactions of the antiparticles that will be introduced in a way that is consistent with the particle-particle interactions from which they are derived. The parameterization of the problem has considerable influence on our intuitive feel for the descriptions, as is demonstrated by the fact that we found it necessary to choose certain parameterizations in order to properly embed interactions in a cluster decomposable way [1-5]. This is why we choose a parameterization in terms of invariant energy (M) and angular orientation parameters ( $\xi = \hat{q}'\cdot\hat{q}$ ) in the standard state reference system (rest frame) for the scattering particles.

The fact that we need an *off-diagonal* and *off-shell* description of the input transition matrix for the general formulation of Faddeev-type integral equations in the relativistic problem means that our variables can only be required to reduce to the usual Mandelstam parameters on-diagonal and on-shell. This is straightforward to accomplish. What gives us more concern is to demonstrate that one is compelled to go off-diagonal using a particular prescription, which will be presented in the following sections.

A. Invariant energy-angle parameters and their relationships to Mandelstam variables

We begin by considering a general two particle in – two particle out process, where the *in* parameters will be on the right (R), and the *out* parameters on the left (L). Diagrammatically, this particle - particle channel (or pp channel) will be represented as follows:

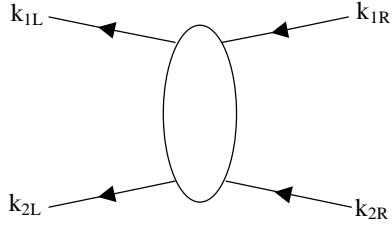

The following parameters are defined in terms of the four-momenta of the particles

$$s_{\frac{L}{R}} \equiv (\vec{k}_{1\frac{L}{R}} + \vec{k}_{2\frac{L}{R}})^2$$
$$t_a \equiv (\vec{k}_{aR} - \vec{k}_{aL})^2$$
$$u_a \equiv (\vec{k}_{aR} - \vec{k}_{\neg aL})^2$$

2.1

where $a \in 1,2$ and the symbol $\neg a$ means the particle other than particle a. Generally, the scattering amplitude need not be on-diagonal; therefore the left-right distinction will be maintained for the present. This will allow the definition of general off-diagonal parameters for the scattering theory. We maintain the following conventions when changing the overall sign of a particle's four-momentum:

$$m_a = \sqrt{\vec{k}_a \cdot \vec{k}_a} \qquad -m_a = \sqrt{-\vec{k}_a \cdot -\vec{k}_a}$$
$$\xi_{t_a} \equiv \hat{k}_{aR} \cdot \hat{k}_{aL} \qquad \xi_{u_a} \equiv -\hat{k}_{aR} \cdot \hat{k}_{\neg aL}$$

2.2

In terms of the energies and momenta of the total system, the four-momenta of each particle can be expressed for either L or R case as

$$\vec{k}_a = \left(\varepsilon(M, m_a, m_{\neg a}), q(M^2, m_a, m_{\neg a})\hat{k}_a\right)$$

**2. 3**

where

$$\varepsilon(M, m_1, m_2) \equiv \frac{1}{2M}(M^2 + m_1^2 - m_2^2)$$

$$q^2(M^2, m_1, m_2) \equiv \frac{[M^2 - (m_1 + m_2)^2][M^2 - (m_1 - m_2)^2]}{4M^2}$$

**2. 4**

For brevity of notation, we will sometimes write these energies and momenta of the particles as

$$\varepsilon_{a_R^L} \equiv \varepsilon(M_{_R^L}, m_{a_R^L}, m_{\neg a_R^L})$$

$$q_{_R^L} \equiv q(M_{_R^L}^2, m_{a_R^L}, m_{\neg a_R^L})$$

**2. 5**

This allows us to write the general off-diagonal forms of the parameters given in equation 2. 1 in terms of the physical invariant energy and angle parameters:

$$s_{_R^L} = M_{_R^L}^2$$

$$t_a = m_{aR}^2 + m_{aL}^2 - 2\varepsilon_{aR}\varepsilon_{aL} + 2q_R q_L \xi_{t_a}$$

$$u_a = m_{aR}^2 + m_{\bar{a}L}^2 - 2\varepsilon_{aR}\varepsilon_{\bar{a}L} - 2q_R q_L \xi_{t_a}$$

**2. 6**

In particular, one can examine these variables when all (energy) parameters are on diagonal, which will be the case for the physical amplitudes. These parameters take on the following form in this limit:

$$M_L = M_R = M \qquad \xi_{t_a} = \xi_{u_a} = \xi$$

$$s_L = s_R = s \qquad t_a = t \qquad u_a = u$$

$$s + t + u = m_{1R}^2 + m_{1L}^2 + m_{2R}^2 + m_{2L}^2$$

**2. 7**

which defines the usual on-diagonal Mandelstam variables *s, t,* and *u.*

We will be particularly interested in the particle-antiparticle symmetry properties of scattering amplitudes. For the problems which will be considered here, there will be conserved particle quantum numbers, and in particular $m_{aL}=m_{aR}$, i.e., the antiparticles generated from the identifications of the symmetry transformation will have the same mass as the corresponding particles.

$$s = M^2$$
$$t = -2q^2(M^2, m_1, m_2)(1-\xi)$$
$$u = \frac{(m_1^2 - m_2^2)^2}{M^2} - 2q^2(M^2, m_1, m_2)(1+\xi)$$

2. 8

These Mandelstam variables have the following threshold behaviors in the pp channel, assuming equality of the right and left masses of a given particle:

$$s \geq (m_1 + m_2)^2$$
$$t \leq 0$$
$$u \leq (m_1 - m_2)^2$$

2. 9

B. Diagrammatic identifications of natural parameters

Next we introduce antiparticles into the two-particle scattering theory by identifying an outgoing positive four-momentum particle with an incoming antiparticle corresponding to identifying the negative of the particle's four-momentum (and vice versa). In particular, we will exchange incoming particle 2 with outgoing particle 2. Diagrammatically, this particle-antiparticle scattering channel (p-pbar or bar channel) is represented below:

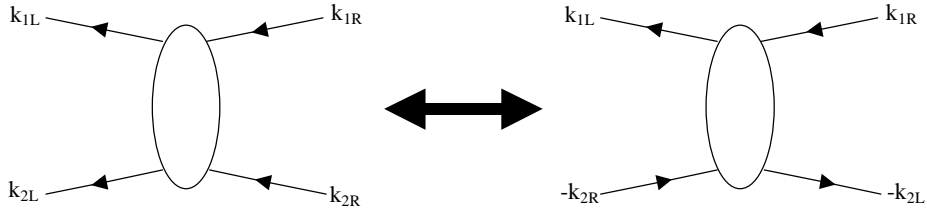

In each diagram, the particles on the left are considered to be kinematically outgoing, and those on the right are incoming. One needs to be able to uniquely define kinematic parameters which map into the physical invariant energies and angles, since the identification shown will map into different reference frames in the pp channel versus the bar channel. However, one can use the momenta of particle 1 which appears in BOTH identifications to uniquely define these parameters:

$$\vec{k}_{1R} = (\varepsilon_{1R}, q_R \hat{k}_{1R}) \qquad \vec{\bar{k}}_{1R} = (\bar{\varepsilon}_{1R}, \bar{q}_R \hat{\bar{k}}_{1R})$$

$$\vec{k}_{1L} = (\varepsilon_{1L}, q_L \hat{k}_{1L}) \qquad \vec{\bar{k}}_{1L} = (\bar{\varepsilon}_{1L}, \bar{q}_L \hat{\bar{k}}_{1L})$$

$$\bar{\varepsilon}_{\substack{L \\ R}} = \frac{1}{2\bar{M}_{\substack{L \\ R}}} \left( \bar{M}_{\substack{L \\ R}}^2 + m_1^2 - (-m_2)^2 \right)$$

$$\bar{q}_{\substack{L \\ R}}^2 = \frac{[\bar{M}_{\substack{L \\ R}}^2 - (m_1 - m_2)^2][\bar{M}_{\substack{L \\ R}}^2 - (m_1 + m_2)^2]}{4\bar{M}_{\substack{L \\ R}}^2}$$

**2. 10**

This allows us to write well defined off-diagonal forms for the invariant energies and angles involved using the two following equations:

$$(\varepsilon_{1R} - \varepsilon_{1L})^2 - (q_R \hat{k}_{1R} - q_L \hat{k}_{1L})^2 = (\overline{\varepsilon}_{1R} - \overline{\varepsilon}_{1L})^2 - (\overline{q}_R \hat{\overline{k}}_{1R} - \overline{q}_L \hat{\overline{k}}_{1L})^2$$

$$\overline{M}_{\substack{L \\ R}}^2 = (\varepsilon_{1_R^L} - \varepsilon_{2_L^R})^2 - (q_{_R^L} \hat{k}_{1_R^L} - q_{_L^R} \hat{k}_{2_L^R})^2 = (\vec{k}_{1_R^L} - \vec{k}_{2_L^R})^2$$

2. 11

The parameters describing the physical amplitudes on-diagonal ($M_L = M_R = M$, etc) will satisfy

$$-2q^2(M^2, m_1, m_2)(1-\xi) = -2q^2(\overline{M}^2, m_1, -m_2)(1-\overline{\xi})$$

$$\overline{M}^2(M^2, \xi) = \left(\frac{m_1^2 - m_2^2}{M}\right)^2 - 2q^2(M^2, m_1, m_2)(1+\xi)$$

2. 12

For general masses $m_1 \neq m_2$, the kinematic ranges of the on-diagonal variables are given by

$$-1 \leq \xi \leq +1 \qquad (m_1 + m_2) \leq M^2 \leq \infty$$
$$-1 \leq \overline{\xi} \leq +1 \qquad (m_1 - m_2)^2 \geq \overline{M}^2 \geq -\infty$$

2. 13

In terms of the usual Mandelstam variables, the following identifications can be made:

$$\overline{s} = \overline{M}^2$$
$$\overline{t} = -2q^2(\overline{M}^2, m_1, -m_2)(1-\overline{\xi})$$
$$\overline{u} = 2(m_1^2 + m_2^2) - \overline{M}^2 - \overline{t}$$
$$\overline{\xi} = 1 + \frac{2(m_1^2 + m_2^2) - \overline{s} + (\overline{t} - \overline{u})}{4q^2(\overline{s}, m_1, -m_2)} = \frac{(m_1^2 - m_2^2)^2}{4\overline{s}q^2(\overline{s}, m_1, -m_2)} + \frac{\overline{t} - \overline{u}}{4q^2(\overline{s}, m_1, -m_2)}$$

2. 14

The final identification will be made by exchanging the outgoing particle 1 with the incoming particle 2. The physical transformation channel (X channel) represents a process which is fundamentally distinct from the prior physical processes, since particle quantum numbers are pairwise annihilated and created. This is represented in the following diagram:

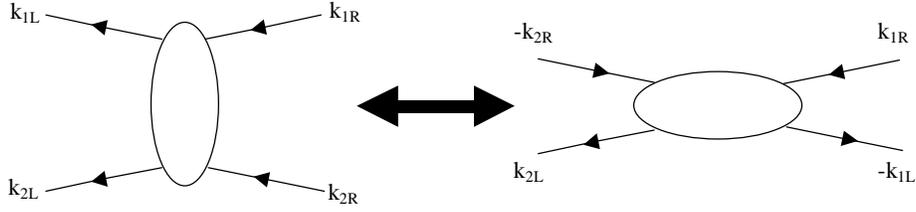

Again, one needs to be able to uniquely define kinematic parameters which map into the physical invariant energies and angles, since the identification shown will map into different reference frames in the pp channel versus the X channel. However, one can use the incoming momentum of particle 1 and outgoing momentum of particle 2 which appears in BOTH identifications to uniquely define these parameters:

$$\vec{k}_{1R} = (\varepsilon_{1R}, q_R \hat{k}_{1R}) \qquad \vec{k}_{1RX} = (\varepsilon_{1RX}, q_{RX} \hat{k}_{1RX})$$
$$\vec{k}_{2L} = (\varepsilon_{2L}, -q_L \hat{k}_{1L}) \qquad \vec{k}_{2LX} = (\varepsilon_{2LX}, -q_{LX} \hat{k}_{1LX})$$
$$\varepsilon_{a^L_R X} = \frac{M_{^L_R X}}{2}$$
$$q_{a^L_R X}^2 = \frac{[M_{^L_R X}^2 - 4m_a^2]}{4}$$

2. 15

This again allows us to write well defined off-diagonal forms for the invariant energies and angles involved using the two equations

$$(\varepsilon_{1R} - \varepsilon_{2L})^2 - (q_R \hat{k}_{1R} + q_L \hat{k}_{1L})^2 = (\varepsilon_{1RX} - \varepsilon_{2LX})^2 - (q_{RX} \hat{k}_{1RX} + q_{LX} \hat{k}_{1LX})^2$$
$$M_{aX}^2 = (\varepsilon_{1aR} - \varepsilon_{aL})^2 - (q_R \hat{k}_{aR} - q_L \hat{k}_{aL})^2 = (\vec{k}_{aR} - \vec{k}_{aL})^2$$

2. 16

The parameters describing the physical amplitudes on-diagonal ($M_L = M_R = M$, etc) will satisfy

$$\left(\frac{m_1^2 - m_2^2}{M}\right)^2 - 2q^2(M^2, m_1, m_2)(1+\xi) = \left(\frac{m_1^2 - m_2^2}{M_X}\right)^2 - \left(q_{1X}^2 + q_{2X}^2 + 2q_{1X} q_{2X} \xi_X\right)$$
$$M_X^2(M^2, \xi) = -2q^2(M^2, m_1, m_2)(1-\xi)$$

2. 17

For general masses $m_1 \neq m_2$, the kinematic ranges of the on-diagonal variables which connect the pp channel to the X channel are given by

$$-1 \leq \xi \leq +1 \qquad m_1 + m_2 \leq M \leq \infty$$
$$-1 \geq \xi_X \geq -\infty \qquad 0 \geq M_X^{\,2} \geq -\infty$$

2. 18

In terms of the usual Mandelstam variables, the following identifications can be made:

$$s_X = M_X^2$$
$$u_X = \left(\frac{m_1^2 - m_2^2}{M_X}\right)^2 - \left(q_{1X}^2 + q_{2X}^2 + 2q_{1X}q_{2X}\xi_X\right)$$
$$t_X = 2(m_1^2 + m_2^2) - M_X^2 - u_X$$
$$\xi_X = \frac{(m_1^2 - m_2^2)^2}{2q(s_X, m_1, -m_1)q(s_X, m_2, -m_2)M_X^2} + \frac{t_X - u_X}{4q(s_X, m_1, -m_1)q(s_X, m_2, -m_2)}$$

2. 19

The various forms for the Mandelstam parameters can immediately be seen to correspond to the usual interpretations in the so-called "s, t, and u channels". The identifications are made when the invariant energy and angle parameters $M^2$ and $\xi$ are obtained by inverting the functional forms specified in the previous equations:

$$\begin{array}{lll} s = M^2 & t = M_X^2(M^2, \xi) & u = \overline{M}^2(M^2, \xi) \\ \overline{s} = \overline{M}^2 & \overline{t} = M_X^2(\overline{M}^2, \overline{\xi}) & \overline{u} = M^2(\overline{M}^2, \overline{\xi}) \\ s_X = M_X^2 & t_X = M^2(M_X^2, \xi_X) & u_X = \overline{M}^2(M_X^2, \xi_X) \end{array}$$

2. 20

The key result of this section has been the demonstration of well defined parameterizations of invariant physical variables in totally disparate physical channels by identifying those particles which appear in both channel descriptions. Off-diagonal identifications for scattering amplitudes can then be made in a compelling way using the forms presented.

### C. Equal mass parameterizations

The analytic relationships between the invariant energy-angle parameters are qualitatively different for the special case in which the masses are equal $m_1=m_2=m$. Since this case will be of particular interest in the development of what follows, we will examine the behavior of these parameters in some detail in this section. In this case the momentum parameters in the various channels have the same functional form

$$q^2(M^2) = \frac{M^2 - 4m^2}{4}$$

2. 21

This means that the form of the connection between parameters in the pp channel and the p-pbar channel is given by

$$q^2(M^2)(1-\xi) = q^2(\overline{M}^2)(1-\overline{\xi})$$
$$\overline{M}^2 = -2q^2(M^2)(1+\xi)$$
$$\overline{\xi} = \frac{M^2(3-\xi) - 4m^2(1-\xi)}{M^2(1+\xi) + 4m^2(1-\xi)}$$
$$0 \geq \overline{M}^2 \geq -\infty \qquad \infty \geq \overline{\xi} \geq 1$$

2. 22

Similarly, the form of the connection between parameters in the pp channel and the annihilation (X) channel is given by

$$q^2(M^2)(1+\xi) = q^2(M_X^2)(1+\xi_X)$$
$$M_X^2 = -2q^2(M^2)(1-\xi)$$
$$\xi_X = -\left(\frac{M^2(3+\xi) - 4m^2(1+\xi)}{M^2(1-\xi) + 4m^2(1+\xi)}\right)$$
$$0 \geq M_X^2 \geq -\infty \qquad -1 \geq \xi_X \geq -\infty$$

2. 23

Finally, one can explicitly represent the connection between the parameters in the p-pbar channel and the X channel given by

$$q^2(\overline{M}^2)(1+\overline{\xi}) = q^2(M_X{}^2)(1-\xi_X)$$
$$M_X{}^2 = -2q^2(\overline{M}^2)(1-\overline{\xi})$$
$$\xi_X = +\left(\frac{\overline{M}^2(3+\overline{\xi}) - 4m^2(1+\overline{\xi})}{\overline{M}^2(1-\overline{\xi}) + 4m^2(1+\overline{\xi})}\right)$$
$$0 \geq M_X{}^2 \geq -\infty \qquad 1 \leq \xi_X \leq \infty$$

<p style="text-align:right">2. 24</p>

All relationships are invertible, and the inversions give those relationships that would be expected from just reassigning which initial channel would be associated with particle-particle scattering.

For a fixed invariant energy M, as the angular parameter ξ varies over all physical values, the other channel parameters vary as demonstrated below

$$\xi : -1 \to +1$$
$$\overline{M}^2 : 0 \to -4q^2(M^2)$$
$$\overline{\xi} : \left(\frac{M^2 - 2m^2}{2m^2}\right) \to 1$$
$$M_X^2 : -4q^2(M^2) \to 0$$
$$\xi_X : -1 \to -\left(\frac{M^2 - 2m^2}{2m^2}\right)$$

<p style="text-align:right">2. 25</p>

Alternatively, if one expresses the parameters in terms of p-pbar channel variables, the other channel parameters vary as

$$\overline{\xi} : -1 \to +1$$
$$M^2 : 0 \to -4q^2(\overline{M}^2)$$
$$\xi : \left(\frac{\overline{M}^2 - 2m^2}{2m^2}\right) \to 1$$
$$M_X^2 : -4q^2(\overline{M}^2) \to 0$$
$$\xi_X : +1 \to +\left(\frac{\overline{M}^2 - 2m^2}{2m^2}\right)$$

<p style="text-align:right">2. 26</p>

These relationships will be useful when determining the constraints on the form of amplitudes which can describe scatterings within the various sectors.

III. DESCRIPTION OF SCATTERING AMPLITUDES AND THE IDENTIFICATION OF ANTIPARTICLES

The relationships between the unitary scattering amplitudes which describe particle-particle scattering and particle-antiparticle scattering are the primary issue to be explored. In the present context, the particle-particle scattering amplitudes will be used to <u>define</u> the properties of the antiparticle through its scatterings with particles and other antiparticles. We will not attempt to use analyticity requirements on the on-shell (though perhaps off-diagonal) scattering amplitudes, especially since some of the physically relevant operations (like complex conjugation or absolute value) cannot be represented as analytic functions. Since there is a considerable literature on the analytic S matrix and crossing symmetries [6,7], to avoid confusion, we will specify the form of the symmetric identification of scattering amplitudes to define antiparticle properties presented here as a symmetric-bar identification, or *symbar* for short. In order to correctly prescribe the properties of the scattering amplitudes, close attention will be paid to the crossing properties of the highly successful model of quantum electrodynamics, as well as unitarity conditions in fixed particle number scattering theory.

A. Cross sections and Bound states

The normalization conventions used here will be chosen for close identification to fixed particle number relativistic scattering theory which directly correspond to non-relativistic scattering amplitudes in the appropriate limits. The on-shell scattering operator is connected to the scattering amplitude through the formal relationship:

$$S(\vec{P}_{(o)}) = 1 + 2\pi i \delta^4(\vec{P} - \vec{P}_{(o)}) A(\vec{P}_{(o)})$$

3.1

The scattering amplitude A is generated from the transition matrix operator T which describes the scattering process, or from a fully renormalized physically unitary amplitude obtained by summing over all (or appropriate) diagrams in a perturbative approach. For the two-particle scattering being considered here the amplitude A is essentially the same as the on-shell transition matrix.

$$\delta(M'-M) T(\underline{k}'_1, \underline{k}'_2 | \underline{k}_1, \underline{k}_2; M = \varepsilon_1 + \varepsilon_2 + i0^+) = \delta^4(\vec{k}'_1 + \vec{k}'_2 - \vec{k}_1 - \vec{k}_2) A(M, \hat{q}' \cdot \hat{q})$$

3.2

where the parameters M and $\hat{q}$ are the invariant energy and direction parameters previously defined. Examining the dimensions of the basis states, this amplitude has dimensions of inverse mass squared = length squared.

The unitarity condition for scattering using our normalization is expressed as follows:

$$T(\underline{k}_1, \underline{k}_2 | \underline{k}_{1o}, \underline{k}_{2o}; Z_1) - T(\underline{k}_1, \underline{k}_2 | \underline{k}_{1o}, \underline{k}_{2o}; Z_2) =$$
$$(Z_2 - Z_1) \int \frac{d^3 k'_1}{\varepsilon'_1 / m_1} \frac{d^3 k'_2}{\varepsilon'_2 / m_2} T(\underline{k}_1, \underline{k}_2 | \underline{k}'_1, \underline{k}'_2; Z_1) \left( \frac{1}{M'-Z_1} \right) \left( \frac{1}{M'-Z_2} \right) T(\underline{k}'_1, \underline{k}'_2 | \underline{k}_{1o}, \underline{k}_{2o}; Z_2)$$

3.3

or in terms of on-shell amplitudes

$$A(M_o, \hat{q} \cdot \hat{q}_o) - A^*(M_o, \hat{q} \cdot \hat{q}_o) =$$
$$-2\pi i \int \frac{d^3 k'_1}{\varepsilon'_1 / m_1} \frac{d^3 k'_2}{\varepsilon'_2 / m_2} A^*(M_o, \hat{q}' \cdot \hat{q}) \delta^4(\vec{k}'_1 + \vec{k}'_2 - M_o \vec{u}_o^{(s)}) A(M_o, \hat{q}' \cdot \hat{q}_o)$$

3.4

The differential cross section is directly expressed in terms of this Lorentz invariant amplitude

$$d\sigma = \prod_a \frac{d^3 k_{af}}{\varepsilon_a / m_a} \frac{(2\pi)^4 \delta^4(\sum \vec{k}_{af} - \vec{k}_{1o} - \vec{k}_{2o}) | A_{fo} |^2}{\left( \sqrt{(\vec{k}_{1o} \cdot \vec{k}_{2o})^2 - m_1^2 m_2^2} \right) / (m_1 m_2)}$$

3.5

where the incident flux factor is defined in terms of the initial state kinematics

$$\frac{\sqrt{(\vec{k}_{1o} \cdot \vec{k}_{2o})^2 - m_1^2 m_2^2}}{m_1 m_2} = \frac{M_o \, q(M_o^2, m_1, m_2)}{m_1 m_2}$$

Defining the phase space factor dΠ given by

$$d\Pi' \equiv \frac{d^3 k'_1}{\varepsilon'_1/m_1} \frac{d^3 k'_2}{\varepsilon'_2/m_2} \delta^4(\vec{k}'_1 + \vec{k}'_2 - M\vec{u}_o^{(s)}) \stackrel{.}{=} \frac{m_1 m_2 \, q(M^2, m_1, m_2)}{M} d^2 \hat{q}'$$

$$\equiv \rho_\pi(M, m_1, m_2) \, d^2 \hat{q}'$$

3. 6

one can immediately express a relationship for the forward scattering unitarity condition

$$\operatorname{Im} A(M, \xi = 1) = -\pi \rho_\pi(M, m_1, m_2) \int_{-1}^{+1} |A(M, \xi)|^2 \, 2\pi \, d\xi$$

3. 7

and the total cross section satisfies the optical theorem in the form

$$\sigma_{Total}(M) = \frac{(2\pi)^4}{\pi} \frac{m_1 m_2}{\sqrt{(\vec{k}_{1o} \cdot \vec{k}_{2o})^2 - m_1^2 m_2^2}} \operatorname{Im} A(M, \xi = 1)$$

3. 8

Care has been taken to explicitly display the parametric dependencies in terms of initial state or final state variables.

Much of the intuitive appeal of this type of approach to examining scattering processes is due to its direct connection to non-relativistic scattering ideas. For completeness, the connection of these amplitudes to the outgoing wave function scattering amplitude f(M,θ) and phase shift parameters $\delta^J(M)$ will be demonstrated. The differential cross section is represented as the modulus squared of the wave amplitude *f*, which can be expressed in terms of the invariant amplitude *A* by

$$f(M, \theta) = (2\pi)^2 \frac{m_1 m_2}{M} A(M, \cos\theta)$$

3. 9

The dimensional factors in this relationship arise from the (on-shell) ratio of the outgoing phase space to the incident flux factors. Similarly, since the invariant amplitude depends only upon the relative angle between incident and final direction parameters, it can be expanded in partial waves

$$A(M, \hat{q} \cdot \hat{q}_o) = \sum_J \frac{2J+1}{4\pi} A^J(M) P_J(\hat{q} \cdot \hat{q}_o) = \sum_{J, J_z} Y_J^{J_z}(\hat{q}) A^J(M) Y_J^{*J_z}(\hat{q}_o)$$

3. 10

which can be used to define phase shifts and absorption parameters for the elastic amplitude

$$\eta^J e^{2i\delta^J} = 1 + 2\pi i \, \rho_\pi(M) A^J(M)$$

3. 11

These various forms of the on-shell amplitude will all satisfy appropriate unitarity conditions as long as the transition amplitude satisfies a unitarity condition.

One can formally examine the bound states of a system directly from the behavior of the transition amplitude off-shell in terms of the Z parameter. On very general grounds, one can expand the eigenstates in the discrete spectrum of an interacting system in terms of the complete set of (usually continuous) states of another system, here chosen to be the asymptotic forms of the (non) scattering states. In the standard state of the pair, the useful result connecting representations of the off-shell $T(Z)$ to sums over eigenstates can be obtained from the formal equation (the Lippman-Schwinger equation [9, 10] or other approaches)

$$T(Z) = (H - H_o) - (H - H_o) \frac{1}{H - Z} (H - H_o)$$

3. 12

By expanding using the basis of eigenstates of the system to be solved, the transition amplitude is seen to have poles at the appropriate discrete eigenvalues of the energy. The invariant transition amplitudes are best examined using a form of the quantum mechanics which preserves the Lorentz frame of the initial and final state off-diagonal, which guarantees that the invariant wave-functions are generated in the same Lorentz system [1]. The Jacobian which transforms the relativistic system to these coordinates is given by

$$\frac{d^3k_1}{\varepsilon_1/m_1}\frac{d^3k_2}{\varepsilon_2/m_2}u^0\delta^3(\underline{u}-\underline{u}_o)=m_1m_2q(M^2,m_1,m_2)M^2\,dM\,\frac{d^3u}{u^0}d^2\hat{q}u^0\delta^3(\underline{u}-\underline{u}_o)$$

$$=\rho_\pi(M,m_1,m_2)\,M^3\,dM\,\frac{d^3u}{u^0}d^2\hat{q}u^0\delta^3(\underline{u}-\underline{u}_o)$$

The off-shell transition amplitude is seen to always have the following form near a bound state:

$$\lim_{Z\to\mu_n}\left[(Z-\mu_n)T(\underline{k}_1,\underline{k}_2\,|\,\underline{k}_{1o},\underline{k}_{2o};Z_1)\right]=$$

$$u^0\delta^3(\underline{u}-\underline{u}_o)\frac{(M-\mu_n)\Psi_{\mu_n\ell\ell_z}(M)Y_\ell^{\ell_z}(\hat{q})}{\sqrt{M^3\rho_\pi(M,m_1,m_2)}}\frac{Y_\ell^{*\ell_z}(\hat{q}_o)\Psi^*_{\mu_n\ell\ell_z}(M_o)(M_o-\mu_n)}{\sqrt{M_o^3\rho_\pi(M_o,m_1,m_2)}}$$

3. 13

where the wave functions are assumed to be normalized according to the condition

$$\int|\Psi_{\mu_n\ell\ell_z}(M)|^2\,dM=1$$

3. 14

Thus, the off-shell behavior of the transition amplitude near a particular bound state pole factorizes into a form which is determined by the (energy-momentum space) wave function of that bound state.

B. Identification of amplitudes in particle-particle, particle-antiparticle, and annihilation-re-creation channels

The primary purpose of the work presented here will be to utilize the properties of one's understanding of formal scattering theory that are applicable when particle type is conserved in cases for which pair creation or annihilation occurs. We are motivated to use the physical restriction that particles and antiparticles can be created or annihilated ONLY pair-wise to <u>define</u> the corresponding properties of antiparticles (or conversely of the particles) through the particle-particle scattering amplitudes. The relationships between the scattering amplitudes of different physical properties between the involved particles and their corresponding antiparticles is typically expressed through the behavior of the amplitude under crossing. The usual identifications in terms of a Lorentz invariant amplitude of the form $\mathcal{M}$ are made as follows:

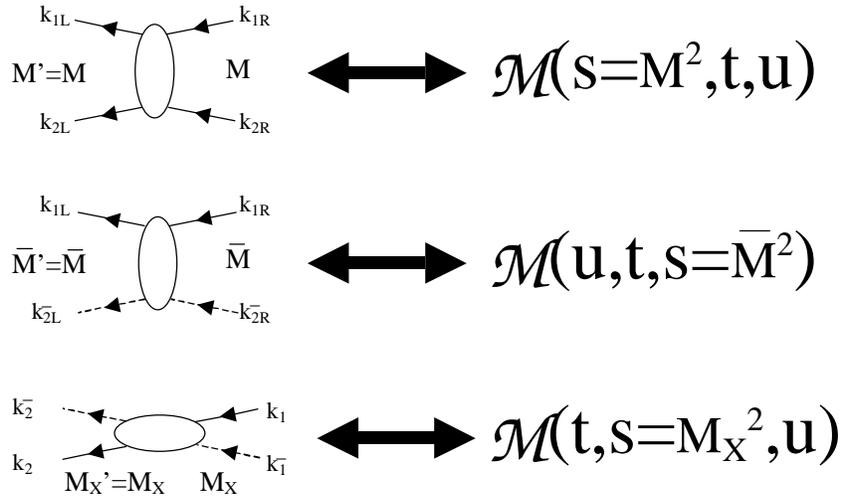

The dashed lines indicate physical antiparticles, and the parameter M is the on-diagonal invariant physical energy for the scattering. We therefore make the following identifications for our (on-diagonal) amplitudes:

$$A(M,\xi) = \mathcal{M}\left(s = M^2, t = M_X^2(M^2,\xi), u = \overline{M}^2(M^2,\xi)\right)$$
$$\overline{A}(\overline{M},\overline{\xi}) = \mathcal{M}\left(\overline{u} = M^2(\overline{M}^2,\overline{\xi}), \overline{t} = M_X^2(\overline{M}^2,\overline{\xi}), \overline{s} = \overline{M}^2\right)$$
$$A_X(M_X,\xi_X) = \mathcal{M}\left(t_X = M^2(M_X^2,\xi_X), s_X = M_X^2, u_X = \overline{M}^2(M_X^2,\xi_X)\right)$$

3. 15

The primary task is to identify these amplitudes as defined with physical two-particle scattering processes in a physically meaningful manner. These amplitudes will be used to define the antiparticle relative to the scattering behavior of its corresponding particle. Note that equation 3.15 establishes for us the *form invariance* of our amplitudes under our anti-particle identification and replacement, thus performing for us the same function that "*s t u* crossing" does in the conventional approach. The unitarity properties of the scattering processes so defined by our **symbar** identifications will be explored shortly.

### C. Relationships to QED crossing properties

For completeness, the corresponding lowest order behavior in QED scattering of two charged particles is presented so that the connections are made clearer. For scattering between leptons of masses $m_1$ and $m_2$, the polarization averaged Lorentz invariant matrix element squared in QED is of the following form for the displayed processes, particle-particle scattering, particle-antiparticle scattering, and particle-antiparticle annihilation and creation:

$$|\mathcal{M}(m_1 m_2 \to m_1 m_2)|^2 = \left(\frac{4\pi e^2}{(2\pi)^3}\right)^2 \frac{1}{2t^2}\left[(u-m_1^2-m_2^2)^2 + (s-m_1^2-m_2^2)^2 + 2(m_1^2+m_2^2)t\right]$$

$$|\mathcal{M}(m_1 m_{\bar{2}} \to m_1 m_{\bar{2}})|^2 = \left(\frac{4\pi e^2}{(2\pi)^3}\right)^2 \frac{1}{2\bar{t}^2}\left[(\bar{s}-m_1^2-m_2^2)^2 + (\bar{u}-m_1^2-m_2^2)^2 + 2(m_1^2+m_2^2)\bar{t}\right]$$

$$|\mathcal{M}(m_1 m_{\bar{1}} \to m_{\bar{2}} m_2)|^2 = \left(\frac{4\pi e^2}{(2\pi)^3}\right)^2 \frac{1}{2s_X^2}\left[(u_X-m_1^2-m_2^2)^2 + (t_X-m_1^2-m_2^2)^2 + 2(m_1^2+m_2^2)s_X\right]$$

For scalar particle scattering through exchange of a single photon, the form is given by

$$|\mathcal{M}(m_1 m_2 \to m_1 m_2)|^2 = \left(\frac{4\pi e^2}{(2\pi)^3}\right)^2 \frac{1}{4t^2}\left[(s-u)^2\right]$$

$$|\mathcal{M}(m_1 m_{\bar{2}} \to m_1 m_{\bar{2}})|^2 = \left(\frac{4\pi e^2}{(2\pi)^3}\right)^2 \frac{1}{4\bar{t}^2}\left[(\bar{u}-\bar{s})^2\right]$$

$$|\mathcal{M}(m_1 m_{\bar{1}} \to m_{\bar{2}} m_2)|^2 = \left(\frac{4\pi e^2}{(2\pi)^3}\right)^2 \frac{1}{4s_X^2}\left[(t_X-u_X)^2\right]$$

3.16

These amplitudes would clearly represent the lowest order behavior of unitary amplitudes as defined by equation [3. 15]. Higher order terms, when properly renormalized and written in terms of physical parameters, are expected to satisfy the same behaviors in terms of the external kinematic variables.

## IV.  GENERAL DEMONSTRATION OF NON-PERTURBATIVE UNITARITY

### A.  Unitarity requirements

The physical unitarity of the scattering amplitudes guarantees probability conservation for kinematically relevant regimes within the elastic two-particle sector, and non-trivially connects the amplitudes for scatterings which couple to inelastic channels. There is little hope of generating a unitary scheme which can include pair creation if the amplitudes which generate these processes are not of themselves unitary below production threshold. Our approach will be guided by noting the complete unitarity maintained by the coupled few-particle channel approach [8,1] as production thresholds are traversed. For instance, in a relativistic three-particle system for which the initial state is a bound pair scattering with a third particle, the elastic scattering amplitude for the initial state smoothly maintains the correct unitary relationship to the total cross section as the available invariant energy begins to allow pair rearrangement and breakup [2]. This unitarity is maintained because the input (2-particle) scattering amplitudes are unitary for all energies, and the coherent coupling of the channels is done in a way which uses the unitarity of the input amplitudes to guarantee the unitarity of the amplitude from which all physical processes can be extracted. The extraction of physically meaningful amplitudes is straightforward and well-defined in that formalism.

We will therefore require that the appropriate amplitudes for distinguishable particle-particle and particle-antiparticle scattering that are obtained from our formulations should maintain unitarity for all energies, and that the unitarity of physical amplitudes which will involve a change in particle number should follow from the unitarity of the amplitudes developed here. Since the two particles for the present are considered to be distinguishable (even for equal masses), one would expect the particle-particle (pp) amplitude to be unitary, as well as the particle-antiparticle (bar) amplitude. However, the transformation (X) amplitude couples differing channels, and is not expected to be unitary of itself, since it represents an off-diagonal element in the overall amplitude. For instance, for two non-identical particles a and b, our identification and replacement construction produces the transformation (X) amplitude $a\bar{a} \leftrightarrow b\bar{b}$, as we discuss in more detail below. This is clear from the lack of an identity term in any sense in the transformation amplitude. However, it is clear that the transformation amplitude CAN couple amplitudes between differing particle types in particle-antiparticle scattering, if one is above production threshold for the more massive pair. Therefore, this amplitude will be incorporated by identifying it as the coupling between channels of a multi-channel scattering amplitude.

The forward scattering unitarity condition on the pp and bar amplitudes is given in equation [3. 7], and is here displayed in terms of the explicit variables involved:

$$\text{Im}\, A(M, \xi = 1) = -\pi \frac{m_1 m_2 q(M^2, m_1, m_2)}{M} \int_{-1}^{+1} |A(M,\xi)|^2\, 2\pi d\xi$$

$$\text{Im}\, \bar{A}(\bar{M}, \bar{\xi} = 1) = -\pi \frac{m_1 m_{\bar{2}} q(\bar{M}^2, m_1, m_{\bar{2}})}{\bar{M}} \int_{-1}^{+1} |\bar{A}(\bar{M},\bar{\xi})|^2\, 2\pi d\bar{\xi}$$

4. 1

From equation 3. 15 which relates each of these amplitudes to that in the annihilation amplitude, and by examining the equation 2. 20 which relate the parameters in terms of each other, each of these expressions is seen to be equivalent to a single connection to the transformation amplitude of the form

$$A(M,\xi) = A_X(M_X(M^2,\xi), \xi_X(M^2,\xi))$$
$$\bar{A}(\bar{M},\bar{\xi}) = A_X(M_X(\bar{M}^2,\bar{\xi}), \xi_X(\bar{M}^2,\bar{\xi}))$$

4. 2

The unitarity conditions place constraints on the functional form of the transformation amplitude. The form of the constraints can be seen by examining the forward unitarity condition expressed in the form

$$\text{Im}\, A(M, \xi = 1) = -\pi \frac{m_1 m_2 q(M^2, m_1, m_2)}{M} \int_{-1}^{+1} |A(M',\xi')|^2\, \delta(M'^2 - M^2) dM'^2\, 2\pi d\xi'$$

4. 3

By direct substitution of the functional forms, this means that

$$\text{Im } A_X(M_X = 0, \xi_X(M^2,1)) =$$

$$-\pi \frac{m_1 m_2 q(M^2, m_1, m_2)}{M} \int |A_X(M'_X, \xi'_X)|^2 \delta(M'^2(M'^2_X, \xi'_X) - M^2) dM'^2 2\pi d\xi$$

4. 4

One is therefore motivated to evaluate the variable change that will result in a final integration over the parameter $M_X$. The phase space factor makes this convenient if one wishes to perform this variable change in terms of the momentum states. However, we will evaluate variables in the chosen representation by evaluating

$$J(M^2) \equiv \frac{m_1 m_2 q(M^2, m_1, m_2)}{M} \int \delta(M'^2(M_X'^2, \xi_X') - M^2) \frac{\partial(M'^2, \xi')}{\partial(M_X'^2, \xi_X')} d\xi_X'$$

$$= \frac{m_1 m_2 q(M^2, m_1, m_2)}{M} \left[\frac{\partial M^2(M_X'^2, \xi_X)}{\partial \xi_X}\right]^{-1} \frac{\partial(M^2, \xi')}{\partial(M_X'^2, \xi_X)}$$

$$= -\frac{m_1 m_2}{2M q(M^2, m_1, m_2)}$$

4. 5

Then the transformation amplitude must satisfy

$$\text{Im } A_X(M_X = 0, \xi_X(M^2,1)) =$$

$$-J(M^2) 2\pi^2 \int_{-4q^2(M^2,m_1,m_2)}^{0} |A_X(M'_X, \tilde{\xi}_X(M_X'^2, M^2))|^2 \, dM_X'^2$$

4. 6

where

$$\tilde{\xi}_X(M_X'^2, M^2) = \frac{\left(\frac{m_1^2 - m_2^2}{M'_X}\right)^2 - \left(\frac{m_1^2 - m_2^2}{M}\right)^2 + 4q^2(M^2, m_1, m_2) - (q_{1X}^2(M_X'^2) + q_{2X}^2(M_X'^2))}{2 q_{1X}(M_X'^2) q_{2X}(M_X'^2)}$$

4. 7

Because of the direct symmetry in the connections between the parameters defined in the transformation channel with those defined in the particle-particle and particle-antiparticle channels, if this relation is satisfied, both the pp and the bar channel will satisfy the optical theorem. A similar set of steps can be followed to establish the unitarity for all angles.

The general form of the constraint equation for arbitrary angles can be expressed using equation 3.4 and the form invariance guaranteed by equation 3.15. The invariant functional form is required to satisfy a unitarity condition for the pp or the ppbar channel given by

$$A(M,\hat{q}\cdot\hat{q}_o) - A^*(M,\hat{q}\cdot\hat{q}_o) = -2\pi i \frac{m_1 m_2 q(M^2,m_1,m_2)}{M} \int d^2\hat{q}' A^*(M,\hat{q}'\cdot\hat{q})A(M,\hat{q}'\cdot\hat{q}_o)$$

$$\overline{A}(\overline{M},\hat{\bar{q}}\cdot\hat{\bar{q}}_o) - \overline{A}^*(M,\hat{\bar{q}}\cdot\hat{\bar{q}}_o) = -2\pi i \frac{m_1 m_2 q(\overline{M}^2,m_1,m_2)}{\overline{M}} \int d^2\hat{\bar{q}}' \overline{A}^*(\overline{M},\hat{\bar{q}}'\cdot\hat{\bar{q}})\overline{A}(\overline{M},\hat{\bar{q}}'\cdot\hat{\bar{q}}_o)$$

4.8

As for the forward scattering case, the integral can be expressed in terms of an integration over the parameter $M_X^2$ using the Jacobian which will properly transform the parameters between the channels given in section II. Substitution of the forms given in equation 4.2 allows us to formally represent these unitarity conditions by rewriting the above equations in the following form:

$$A_X(M_X(M^2,\hat{q}\cdot\hat{q}_o),\xi_X(M^2,\hat{q}\cdot\hat{q}_o)) - A_X^*(M_X(M^2,\hat{q}\cdot\hat{q}_o),\xi_X(M^2,\hat{q}\cdot\hat{q}_o)) =$$
$$-2\pi i \frac{m_1 m_2 q(M^2,m_1,m_2)}{M} \int \int \delta(M'^2 - M^2) \frac{\partial(M'^2,\hat{q}')}{\partial(M_X'^2,\hat{q}_X')} dM_X'^2 d^2\hat{q}_X' \otimes$$
$$A_X^*(M_X(M^2,\hat{q}'\cdot\hat{q}),\xi_X(M^2,\hat{q}'\cdot\hat{q})) A_X(M_X(M^2,\hat{q}'\cdot\hat{q}_o),\xi_X(M^2,\hat{q}'\cdot\hat{q}_o))$$

4.9

Alternatively, this expression can be written in terms of the bar channel kinematic parameters, demonstrating the form invariance of the unitarity constraint under our replacement and substitution process (symbar). This is the general formal result we wish to demonstrate for the pp and p-pbar unitarity condition needed for the construction of a finite particle number scattering formalism.

A general property of the Jacobian of transformation between independent variables is the chain rule

$$\frac{\partial(M^2,\hat{q})}{\partial(\overline{M}^2,\hat{\bar{q}})} = \frac{\partial(M^2,\hat{q})}{\partial(M_X^2,\hat{q}_X)} \frac{\partial(M_X^2,\hat{q}_X)}{\partial(\overline{M}^2,\hat{\bar{q}})}$$

which connects one coordinate transformation to another. This means the one can directly relate the unitarity conditions of the pp channel and the p-pbar channel without expressing the amplitudes in terms of the X channel if desired. The form of the kinematic transformation between the variables representing the different channels as defined in section II can be seen to be symmetrical in form (up to sign differences in the angular parameters) in the relationship of the pp channel and the p-pbar channel to the X channel variables. Although in practice, for unequal masses and arbitrary angles it might be difficult to obtain explicit functional forms for the relations between the energy and angular parameters of the different channels, in principle this allows us again to generally express the form of the unitarity condition for either pp channel or p-pbar channel in terms of a single constraint equation on the form of the X channel amplitude.

### B. Multichannel identifications

The most complicated aspect of the unitarity relationships involves the inclusion of particle annihilation in a consistent way. As has been mentioned, the transformation channel is unique in that particle quantum numbers can be pairwise created and annihilated. This means that in order to properly include annihilation in a unitary formulation of scattering theory, this particular channel serves as a coupling into different particle sectors. A unitary coupled channel formulation is particularly well suited for this endeavor. In fixed particle number relativistic few particle dynamics, the unitarity of the full scattering system is guaranteed because of the complete unitarity of the input channels, which couple in a well defined manner [1]. This remains true even when traversing kinematic thresholds which fragment or rearrange the various clusters. We will utilize this method to include mixing between elastic channels which are themselves unitary at all energies in the absence of the couplings.

For the purpose of clarity of presentation, we will presume that an angular momentum decomposition has been done in the invariant center of momentum reference system, which will expand the amplitudes as partial waves, and allows algebraic demonstrations of the unitarity relationships. The particle-particle scattering amplitude will be represented by the diagram

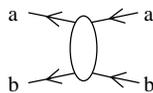

Similarly, the particle-antiparticle (bar channel) amplitude will be represented by the diagram

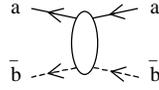

and the transformation (X) channel amplitude will be represented by the diagram

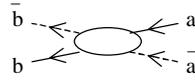

These diagrams are assumed to be directly related to the Lorentz invariant scattering amplitude for a given partial wave, which will have unit modulus when probability flux conservation applies. One would thus expect the pp channel and p-pbar channel amplitudes to individually be unitary, regardless of the presence of other kinematically accessible channels. The issue then becomes one of preserving this form of unitarity when the coupling to other channels coherently interferes with the direct scattering process. For the present purposes, this will only concern us when examining particle-antiparticle scattering, since then the transformation channel will serve as a coupling to other particle states. In what follows we will identify the physical parameterization that will connect the amplitudes generated by particle-antiparticle symmetry operations to the observable scattering processes.

The unitarity condition for multi-channel few particle scattering takes the form

$$\operatorname{Im} A_{b\bar{b}}^{J}(\overline{M}) = \sum_{d} \pi \frac{m_d m_{\bar{d}} q(\overline{M}^2, m_d, m_{\bar{d}})}{M} | A_{db}^{J}(\overline{M}) |^2$$
$$= \sum_{d} \pi \rho_{\pi}(\overline{M}, m_d, m_{\bar{d}}) | A_{db}^{J}(\overline{M}) |^2$$

4. 10

The phase shift δ, inelasticity parameter η, and elastic cross section $\sigma_{elastic}$ get introduced in the elastic channel with the identification

$$\frac{\eta_b^J(\overline{M})e^{2i\delta_b^J(\overline{M})}-1}{2i} \equiv \pi\rho_\pi(\overline{M},m_b,m_{\bar{b}})A_{bb}^J(\overline{M})$$

$$\sigma_{elastic}(\overline{M}) \equiv \sum_J \frac{2J+1}{4\pi}\frac{(2\pi)^4}{\pi}\frac{m_b m_{\bar{b}}}{\left[(\vec{k}_b\cdot\vec{k}_{\bar{b}})^2 - m_b^2 m_{\bar{b}}^2\right]^{\frac{1}{2}}} \pi\rho_\pi(\overline{M},m_b,m_{\bar{b}})|A_{bb}^J(\overline{M})|^2$$

$$= \sum_J (2J+1)4\pi^3 \frac{m_b^4}{\overline{M}^2}|A_{bb}^J(\overline{M})|^2$$

<div align="right">4. 11</div>

Then the unitarity condition defines the inelasticity parameter in terms of the couplings to other channels given by

$$1 - \left(\eta_b^J(\overline{M})\right)^2 = \sum_{d(\neq b)} 2\pi\rho_\pi(\overline{M},m_d,m_{\bar{d}})|A_{db}^J(\overline{M})|^2 2\pi\rho_\pi(\overline{M},m_b,m_{\bar{b}})$$

<div align="right">4. 12</div>

Therefore, the transformation (X) channels which will *define* the inelastic coupling amplitudes changes the normalization of the uncoupled elastic (but unitary) amplitudes through contributions to the inelasticity parameter as given in this equation. Using our normalizations, the total cross section will satisy the optical theorem given in the form

$$\sigma_{Total}(M) = \sum_J \frac{2J+1}{4\pi}\frac{(2\pi)^4}{\pi}\frac{m_o m_{\bar{o}}}{\left[(\vec{k}_o\cdot\vec{k}_{\bar{o}})^2 - m_o^2 m_{\bar{o}}^2\right]^{\frac{1}{2}}} \text{Im}\,A_{oo}^J(M)$$

$$= \sum_J (2J+1)4\pi^2 \frac{m_o^2}{M\,q(M^2,m_o,m_{\bar{o}})} \text{Im}\,A_{oo}^J(M)$$

<div align="right">4. 13</div>

where "o" represents the initial (original) particle-antiparticle pair.

To gain insight into how the "symbar" generated amplitudes will be unitary, we will consider a specific kinematic regime for which the scattering will allow only two particle channels to couple. We will represent the form of the scattering amplitude as

$$S = \begin{pmatrix} \begin{array}{cc} \includegraphics{aa} & \includegraphics{ab} \\ \includegraphics{ba} & \includegraphics{bb} \end{array} \end{pmatrix} e^{i\chi_{ab}}$$

where $\chi_{ab}$ is an overall phase parameter. The diagram of the form will be assumed to be independently unitary in the absence of the coupling. Later, it will be indentified with all coherent amplitudes that contribute to the elastic channel. We parameterize any particular partial wave in terms of the Stapp parameters [11]

$$S = \begin{pmatrix} \cos 2w_{a \atop b} e^{2i\delta_a^S} & i\sin 2w_{a \atop b} e^{i(\delta_a^S + \delta_b^S)} \\ i\sin 2w_{a \atop b} e^{i(\delta_a^S + \delta_b^S)} & \cos 2w_{a \atop b} e^{2i\delta_b^S} \end{pmatrix}$$

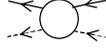

**4. 14**

The identifications of the Stapp parameters with the general coupled channel parameters can be made immediately

$$\eta_a^J(\overline{M}) = \cos 2w_{a \atop b} = \eta_b^J(\overline{M})$$
$$\sin 2w_{a \atop b} = 2\pi \rho_\pi(\overline{M}, m_b, m_{\bar{b}}) | A_{ba}^J(\overline{M}) | \pi \rho_\pi(\overline{M}, m_a, m_{\bar{a}})$$
$$\delta_a^S = \delta_a^J(\overline{M}) \qquad \delta_b^S = \delta_b^J(\overline{M})$$

which demonstrates that the phase parameter defined by Stapp is always identified with the phase shift parameter for coupled channels. In this parameterization, the phase of the transformation amplitude is given in terms of the phase parameters, and its amplitude directly defines the degree of the coupling. If there is zero coupling ($\sin 2w_{a \atop b} = 0$), the two channels are independently unitary and scatter as defined by their elastic phase shifts.

C.     Identical particles and channel interference

For identical particle-antiparticle scattering, we expect coherent interference between the scattering channel and the annihilation channel; for instance, in lowest order QED the electron –positron amplitude generated by single photon exchange interferes with the Bhabha term. To gain insight into how this will affect the amplitudes which are calculated without coherent interference, we will examine first the behavior of the identical channel behavior of the two-channel unitary representation. Diagrammatically this should represent itself as follows:

$$S = \begin{pmatrix} \text{◯}_{+}\text{⨝} & \text{⨝} \\ \text{⨝} & \text{◯}_{+}\text{⨝} \end{pmatrix} e^{i\chi}$$

We will write the total amplitude using the Stapp parameterization given by

$$S = \begin{pmatrix} \cos 2w_T e^{2i\delta_T} & i\sin 2w_T e^{2i\delta_T} \\ i\sin 2w_T e^{2i\delta_T} & \cos 2w_T e^{2i\delta_T} \end{pmatrix} e^{i\chi}$$

We expect the particle-antiparticle scattering channel to itself be unitary. Thus we can write

$$\text{◯} = e^{2i\delta}$$

which allows us to identify the parameters for the scattering amplitude

$$e^{2i\delta} + i\sin 2w_T e^{2i\delta_T} = \cos 2w_T e^{2i\delta_T}$$

**4. 15**

Therefore the total phase parameter is given by

$$\delta_T = \delta + w_T$$

which defines coherent, unitary particle-antiparticle scattering amplitudes in terms of the p-pbar channel and the transformation (X) channel. One can immediately derive these results as the identical particle limit of the distinguishable phase and coupling parameters previously derived

$$\Rightarrow 2\delta_a^S = 2\delta_b^S = 2\delta_T + \chi$$

This discussion illustrates some interesting points. First, if the S matrix given for the indistinguishable limit of the coupled channel representation is to have the same phase as the S matrix which is directly calculated from the amplitudes, the overall phase for the coupled channel representation must satisfy

$$\chi = -2w_T$$

This phase corrects the double counting of the particle transformation processes in this limiting form of the coupled channel S matrix. This result can be obtained directly by noting that for indistinguishable channels, the two-channel eigenstate must not distinguish between the two-channels, which is insured by using an overall 2-channel wave function of the form

$$\Phi = \frac{1}{\sqrt{2}} \begin{pmatrix} 1 \\ 1 \end{pmatrix}$$

It is also interesting to note that the phase shift for the elastic channel adds coherently with the coupling parameter which is determined by the amplitude of the annihilation channel. This matches the predictions of lowest order perturbative expansions, but is obtained non-perturbatively. Finally, a renormalization of the already unitary amplitude for elastic p-pbar scattering occurs due to the coupling. The renormalization is due to the use of distinguishable channel particle wave functions to calculate the original amplitudes, which result in an overall modulus renormalization for the identical channels. The coherent amplitude satisfies the form given in equation 4.15 which can be expressed using the diagram

$$\frac{\text{⟨⟩} + \text{⟨⟩}}{\sqrt{1 - |\text{⟨⟩}|^2}} = e^{2i\delta^T} = \text{⟨⟩}$$

where the equality holds in the absence of coupling to other particle-antiparticle pairs. This renormalization clearly becomes unity when the coupling due to annihilation vanishes.

We can therefore expect a generalization of this behavior for the multichannel description to take the form

$$1 + 2\pi i \rho_\pi (\overline{M}, m, m)\left( \overline{A}^J(\overline{M}) + A_X^J(M_X = \overline{M}) \right) = N^J(\overline{M}) e^{2i\delta_J^T}$$

**4. 16**

If we define the phase shift amplitudes $\tau$ by the formula

$$s^J(\overline{M}) \equiv 1 + 2\pi i \rho_\pi(\overline{M}, m, m) \overline{A}^J(\overline{M}) \equiv e^{2i\delta_J}$$
$$\tau_X^J(\overline{M}) \equiv \pi \rho_\pi(M_X, m, m) A_X^J(M_X) \equiv |\tau_X^J(\overline{M})| e^{2i\chi_J}$$

**4. 17**

then the coherent phase shift and renormalization parameters are given by

$$\tan 2\delta_J^T = \frac{\sin 2\delta_J + |2\tau_X^J|\cos 2\chi_J}{\cos 2\delta_J - |2\tau_X^J|\sin 2\chi_J}$$

$$(N^J)^2 = 1 + |2\tau_X^J|^2 + 2|2\tau_X^J|\sin 2(\delta_J - \chi_J)$$

4. 18

These parameters will give results which are consistent with an order by order expansion of the amplitudes, as well as for the analytically complete results of perturbative sums.

We can also expect to be able to express the coherent amplitude for identical particle-antiparticle scattering to be given in terms of a renormalized sum of the properly symmeterized angular represented amplitudes previously obtained. For the case of scalar particles, the total scattering amplitude will be given by the coherent sum of the symmetric amplitudes

$$\overline{A}_X(\overline{M},\overline{\xi}) = N(\overline{M})\left[A_X^S(M_X(\overline{M},\overline{\xi}),\xi_X(\overline{M},\overline{\xi})) + A_X^S(\overline{M},\overline{\xi})\right]$$

4. 19

The first term is the contribution from the pp-bar amplitude, and the second is the contribution from the transformation (X) amplitude. The renormalization constant is calculated directly from the optical theorem

$$N(\overline{M}) = \frac{\operatorname{Im} A_X^S(0,\xi_X(\overline{M},1)) + \operatorname{Im} A_X^S(\overline{M},\overline{\xi})}{\operatorname{Im} A_X^S(0,\xi_X(\overline{M},1)) + 2\pi\rho_\pi(\overline{M})\int_{-1}^{+1} d\overline{\xi}\left[|A_X^S(\overline{M},\overline{\xi})|^2 + 2\operatorname{Re}\left\{A_X^S(M_X(\overline{M},\overline{\xi}),\xi_X(\overline{M},\overline{\xi}))A_X^{S*}(\overline{M},\overline{\xi})\right\}\right]}$$

4. 20

This form will give results which are consistent with an order by order expansion of the amplitudes, as well as for the analytically complete results of perturbative sums. The renormalization constant can be reexpressed in terms of partial wave amplitudes which satisfy equation 3. 10 by the equation

$$N(\overline{M}) = \frac{\sum_J \left(\frac{2J+1}{4\pi}\right)\left[\operatorname{Im}\overline{A}^J(\overline{M}) + \operatorname{Im}A_X^J(\overline{M})\right]}{\sum_J \left(\frac{2J+1}{4\pi}\right)|\overline{A}^J(\overline{M}) + A_X^J(\overline{M})|^2}$$

4. 21

This gives relationships between the individual partial wave amplitudes and the overall normalization constant for the coherent combination of particle-antiparticle physical scattering amplitudes.

V. EXPLICIT MODELS

We present here two explicit amplitudes which exhibit Lorentz invariance and unitarity, with antiparticles appropriately introduced in the way described in previous sections. The first is a scattering length model acting only in s-waves, chosen for simplicity. To motivate the form of the second model, and make a direct connection to physics familiar to most, the construction will begin from the solution of the non-relativistic Coulomb problem; this example is relevant in many quantum mechanical systems. We will make the model both Lorentz invariant and finite, and then use the model to demonstrate the unitary behavior of the derived amplitudes.

A. An S-wave scattering model for scalar particles

The primary purpose of our presentation has been to demonstrate the particle-antiparticle properties of a unitary fixed particle number scattering formalism. We will therefore construct a minimal model to directly demonstrate the unitarity of all derived amplitudes. In order to utilize the results of the previous sections, the kinematic parameters of the defining amplitude must be expressed in one particular channel. We will choose to express the kinematic content in the parameters of the annihilation channel in order to take advantage of relationships such as equation 4. 2. One only needs to express the momentum $q$ that appears in equation 4. 8 in terms of the parameters $M_X$ and $\xi_X$. There are several model dependent ways for which this can be done. An obvious identification would be to use the relationship generated by the invariant energy on-diagonal delta function which results in the relationship (for equal masses)

$$4q^2 = 2q_X^2(M_X^2)(1+\xi_X) - M_X^2$$
$$4\bar{q}^2 = 2q_X^2(M_X^2)(1-\xi_X) - M_X^2$$

Such an identification will result in many terms in the multipole expansions in the angular dependence of the amplitude in the annihilation channel which will coherently interfere with corresponding terms in the p-pbar channel. Instead, for the first example, we will construct a model that consists of a single scalar (J=0) quantum exchange, so that only one term of the multipole expansion of the particle-antiparticle scattering amplitude will have coherent interference. If there is no angular dependence in the transformation amplitude, it can only depend upon the total integration range over its angle. By examining the range of the parameter $\xi_X$ as given by equations 2. 22, 2. 23, and 2. 24 , we can obtain the following for the various channels

pp channel:

$$\xi_X : -1 \to -\left(\frac{M^2 - 2m^2}{2m^2}\right) \qquad \xi : -1 \to +1$$

$$\delta\xi_X(M,\xi) = -2\frac{q^2}{m^2}$$

p-pbar channel:

$$\xi_X : +1 \to +\left(\frac{\overline{M}^2 - 2m^2}{2m^2}\right) \qquad \overline{\xi} : -1 \to +1$$

$$\delta\xi_X(\overline{M},\overline{\xi}) = 2\frac{\overline{q}^2}{m^2}$$

Transformation (X) channel:

$$\xi_X : -1 \to +1$$
$$\delta\xi_X = 2$$

It is convenient to define the particle type parameter $\zeta_{12}$ for the incoming (or equivalently, outgoing) particles by

$$\zeta_{12} \equiv \frac{m_1 m_2}{|m_1 m_2|} = \begin{cases} +1 & p-p \\ -1 & p-\overline{p} \end{cases}$$

Defining $\delta\xi_X$ to be the integration range of the parameter $\xi_X$ in terms of the appropriate channel variables, the invariant momentum and energy parameters can be defined by

$$Q^2(\delta\xi_X) = -\frac{\zeta_{12}}{2} m^2 \delta\xi_X$$
$$M^2(\delta\xi_X) \equiv 4(m^2 + q^2(\delta\xi_X))$$

Therefore, we will model an s-wave zero range scattering length amplitude using the form

$$A_X(M_X, \delta\xi_X) = \frac{1}{4\pi} \frac{M(\delta\xi_X)}{(2\pi m)^2 Q(\delta\xi_X)} \left( \frac{\sqrt{m^2 - \frac{\mu_Q^2}{4}}}{Q(\delta\xi_X) - i\sqrt{m^2 - \frac{\mu_Q^2}{4}}} \right)$$

5. 1

The corresponding pp channel, p-pbar, and X channel amplitudes will then be given by

$$A(M, \xi) = \frac{M}{(2\pi m)^2 q(M^2)} \left( \frac{-\sqrt{m^2 - \frac{\mu_Q^2}{4}}}{q(M^2) + i\sqrt{m^2 - \frac{\mu_Q^2}{4}}} \right)$$

$$\overline{A}(\overline{M}, \overline{\xi}) = \frac{\overline{M}}{(2\pi m)^2 q(\overline{M}^2)} \left( \frac{\sqrt{m^2 - \frac{\mu_Q^2}{4}}}{q(\overline{M}^2) - i\sqrt{m^2 - \frac{\mu_Q^2}{4}}} \right)$$

$$A_X(M_X, \xi_X) = \frac{2\sqrt{2}}{(2\pi m)^2} \left( \frac{\sqrt{m^2 - \frac{\mu_Q^2}{4}}}{m - i\sqrt{m^2 - \frac{\mu_Q^2}{4}}} \right)$$

5. 2

The chosen form has the following characteristics:

a. The scattering goes only through s-waves in any of the channels

b. There is a single bound state of mass $\mu_Q$ in the p-pbar channel

c. The transformation amplitude is energy independent. However, the transformation coupling will be energy dependent due to the phase space factor.

The unitarity of the pp and p-pbar channel amplitudes can be immediately demonstrated from equation 4.

$$\text{Im} A^J(M) = \pi \frac{m^2 q(M^2, m, m)}{M} |A^J(M)|^2$$

$$\text{Im} \overline{A}^J(\overline{M}) = \pi \frac{m^2 q(\overline{M}^2, m, m)}{\overline{M}} |\overline{A}^J(\overline{M})|^2$$

5. 3

where only the J=0 term is nonvanishing.

We finally demonstrate a unitary form for the coherent amplitude resulting from identical particle-antiparticle elastic scattering and annihilation. An s-wave scattering corresponds to an overall symmetric scattering state. The form demonstrated for the bar amplitude is explicitly unitary in the s-wave, and the transformation channel amplitude has been chosen to only contribute to the s-wave scattering. The coherent superposition of amplitudes will therefore have a unitary form from equation 4. 18 with parameters given by

$$\tan(2\chi_0) = \sqrt{1 - \left(\frac{\mu_Q}{2m}\right)^2}$$

$$|\tau_X^0| = \frac{q(\overline{M}^2)}{\overline{M}} 2\sqrt{2} \frac{\sqrt{1 - \left(\frac{\mu_Q}{2m}\right)^2}}{\sqrt{2 - \left(\frac{\mu_Q}{2m}\right)^2}}$$

$$\tan 2\delta_0 = \frac{2 q(\overline{M}^2)\sqrt{1 - \left(\frac{\mu_Q}{2m}\right)^2}}{q^2(\overline{M}^2) - m^2 + \frac{\mu_Q^2}{4}}$$

5. 4

These parameters then define the overall phase shift $\delta_0^T$ for the identical particle-antiparticle scattering channel which then can couple to other pairs above the appropriate production thresholds.

$$\tan 2\delta_0^T = \frac{\sin 2\delta_0 + |2\tau_X^0|\cos 2\chi_0}{\cos 2\delta_0 - |2\tau_X^0|\sin 2\chi_0}$$

$$(N^0)^2 = 1 + |2\tau_X^0|^2 + 2|2\tau_X^0|\sin 2(\delta_0 - \chi_0)$$

5.5

Because of the phase space factors, there is no coupling to these channels below the kinematic thresholds for pair production, although the amplitudes can exhibit any appropriately unitary behaviors near the thresholds, and the transformation channel amplitude is energy independent. The factor $|\tau_X^0|$ is seen to vanish near production threshold, which decouples the X channel from the elastic channel.

### B. Non-relativistic Coulomb scattering of scalar particles

A form of the solution for the non-relativistic scattering of scalar particles derived from the Shrodinger equation has been given by Mott and Massey [12,13]. For completeness, some aspects of the result is given here, primarily to demonstrate the form utilized by the authors which will be explicitly unitary. For the scattering of a particle of reduced mass $\mu$ and charge $Z_1 e$ from a Coulomb field generated by a fixed charge $Z_2 e$, the wave function takes the form

$$\psi_k(r,\vartheta) = N_k e^{ikz} {}_1F_1(-iZ_1 Z_2 \alpha \frac{\mu}{k}; 1; ik(r-z))$$

5.6

in terms of a confluent hypergeometric function and the fine structure constant $\alpha$. The normalization $N_k$ chosen by Mott and Massey was such that the solution yielded the classical result for Rutherford scattering, and the incoming asymptotic wave form has unit flux near the z-axis. However, due to the long range nature of the Coulomb field, the asymptotic waveforms are considerably distorted, and since there is a forward scattering singularity in the amplitude, the flux normalization cannot be easily checked. In our approach, we will choose this normalization to give unitary results in the *outgoing* flux, which, since the scattering is elastic, will guarantee unitarity in the incoming flux when integrated over the entire incoming distorted hyperbolic wave form.

The asymptotic form of the wave function is most directly obtained by examining an integral representation of the confluent hypergeometric function obtained by taking a closed contour integral over a curve γ which includes the point z and the point t=0 (usually chosen to be a circle around the origin) in the form given by (see [12] for details)

$$_1F_1(a,b;z) = \frac{(b-1)!}{2\pi i}\int_\gamma \left(1-\frac{z}{t}\right)^{-a} e^t t^{-b}\, dt$$

The parts of the contour around the two aforementioned points give a convenient decomposition of the solution into parts which have direct representation for outgoing and incoming asymptotic waveforms:

$$_1F_1(a,b;z) = W_1(a,b;z) + W_2(a,b;z)$$

5.7

The form $W_1$ has an asymptotic form which represents and incoming waveform, and $W_2$ represents and outgoing waveform

$$W_1(a,b;z) \xrightarrow[z\to -\infty]{} \frac{\Gamma(b)}{\Gamma(b-a)}(-z)^{-a} g(a, a-b+1; -z)$$

$$W_2(a,b;z) \xrightarrow[z\to \infty]{} \frac{\Gamma(b)}{\Gamma(a)} e^z z^{a-b} g(1-a, b-a; z)$$

where

$$g(\alpha,\beta;z) \equiv 1 + \frac{\alpha\beta}{z} + \frac{\alpha(\alpha+1)\beta(\beta+1)}{2!z^2} + \cdots$$

We will require that the outgoing flux from the wave function should satisfy a unitarity condition. The function $W_2$ satisfies

$$W_2(-iZ_1Z_2\alpha\frac{\mu}{k};1;ik(r-z)) \xrightarrow[r\to\infty]{} \frac{e^{ik(r-z)}[ik(r-z)]^{-1-iZ_1Z_2\alpha\frac{\mu}{k}}}{\Gamma(-iZ_1Z_2\alpha\frac{\mu}{k})}\left(1 + \frac{\left(1+iZ_1Z_2\alpha\frac{\mu}{k}\right)}{ik(r-z)} + \cdots\right)$$

We will therefore choose the normalization constant to satisfy

$$N_k \equiv \sin\Delta_o(k) 2i\Gamma(iZ_1Z_2\alpha\frac{\mu}{k})e^{-\frac{\pi}{2}Z_1Z_2\alpha\frac{\mu}{k}} e^{iZ_1Z_2\alpha\frac{\mu}{k}\log(2kr)} e^{iv}$$

5.8

This will then insure that the outgoing waveform is given by

$$\psi_k(r,\vartheta) \xrightarrow[r\to\infty]{} \frac{e^{ikr}}{r} f_k(\vartheta)$$

**5. 9**

where

$$f_k(\vartheta) = \frac{\sin \Delta_o(k)}{k} \left(\frac{2}{1-\cos\vartheta}\right)^{1+iZ_1Z_2\alpha\frac{\mu}{k}} \frac{\Gamma(iZ_1Z_2\alpha\frac{\mu}{k})}{\Gamma(-iZ_1Z_2\alpha\frac{\mu}{k})} e^{i\upsilon(k)}$$

**5. 10**

We will shortly explore the unitarity behavior of this function in the model to be constructed. This form is seen to have bound state singularities due to the gamma function in the numerator if the signs of $Z_1$ and $Z_2$ differ, and the argument goes off-shell in such a way that the argument of the gamma function is a negative integer. These singularities are found to exactly correspond to the Bohr spectrum for hydrogenic atoms:

$$k \to i\sqrt{2\mu E_n^B}$$
$$E_n^B = \frac{1}{2}\frac{(Z_1Z_2\alpha)^2 \mu c^2}{n^2}$$

Off-shell extensions of this amplitude will NOT give bound states for charges of the same sign, as would be expected for repulsive interactions.

The incoming waveform will be distorted (hyperbolic) plane waves due to the long range behavior of the Coulomb field. However, if the outgoing flux satisfies a unitarity condition, since the scattering is elastic, the integrated incoming flux will be of a form that will necessarily conserve probability flux. For completeness, the asymptotic form of this incoming waveform is exhibited below:

$$\psi_k(r,\vartheta) \xrightarrow[z\to-\infty]{} \frac{2\sin\Delta_o(k)}{Z_1 Z_2 \alpha \frac{\mu}{k}} e^{iZ_1 Z_2 \alpha \frac{\mu}{k}\log[2kr(r-z)]} e^{iv(k)} \otimes$$

$$e^{ikz}\left[1 + \frac{\left(iZ_1 Z_2 \alpha \frac{\mu}{k}\right)^2}{ik(r-z)} + \cdots\right]$$

5. 11

To construct a demonstrably unitary amplitude for the outgoing flux, we will include a finite quantum mass $m_Q$ in a way which results in the required Coulomb form when $m_Q \to 0$. The finite quantum mass will then provide a cutoff in the range of the Coulomb interaction of the order of the Compton wavelength of the quantum $\frac{\hbar}{m_Q c}$. If one utilizes the form of the invariant energy in the transformation channel

$$M_X^2 = -2q^2(1-\cos\vartheta)$$

then the following form for the amplitude $f_q$ can be directly shown to be satisfy the optical theorem:

$$f_q(\vartheta) = \frac{\sin\Delta(q,\vartheta=0)}{q}\left(\frac{m_Q^2 + 4q^2}{m_Q^2 - M_X^2}\right) e^{i\Delta(q,\vartheta)}$$

5. 12

Note that the form in the parenthesis has been specifically chosen since in the zero quantum mass limit it behaves as follows:

$$\left(\frac{m_Q^2 + 4q^2}{m_Q^2 - M_X^2}\right) \xrightarrow[m_Q \to 0]{} \left(\frac{2}{1-\cos\vartheta}\right)$$

Therefore, we model the Coulomb interaction in a form which makes use of this result.

In the Born limit, one expects to recover the Rutherford scattering result. This constrains the form of the factor $\Delta_o$ in equation 5. 12 to satisfy

$$\sin \Delta_o(q) \underset{BornLimit}{\Rightarrow} \frac{1}{2} Z_1 Z_2 \alpha \frac{\mu}{q}$$

**5. 13**

A direct examination of the previously demonstrated incoming form for the wave function in equation 5. 11 will then have the correct angular dependence to give an appropriate unitary outgoing form, and in the Born limit appear as an incoming plane wave. The overall phase must then be chosen to match the Coulomb form. This phase is given by

$$\Delta(q,\vartheta) = Z_1 Z_2 \alpha \frac{\mu}{q} \log\left(\frac{2}{1-\cos\vartheta}\right) + 2\tilde{\eta}(Z_1 Z_2 \alpha \frac{\mu}{q}) + v(q)$$

**5. 14**

where

$$e^{2i\tilde{\eta}(a)} \equiv \frac{\Gamma(ia)}{\Gamma(-ia)}$$

**5. 15**

and the extra phase factor $v$ will be chosen to have correspondence with Rutherford scattering. A form which will insure this is given by

$$v(q) = \frac{1}{2} Z_1 Z_2 \alpha \frac{\mu}{q}(1 + 4\gamma - 2\pi) - \underset{\substack{\vartheta \to 0 \\ Born}}{Lim} Z_1 Z_2 \alpha \frac{\mu}{q} \log\left(\frac{2}{1-\cos\vartheta}\right)$$

where the factor $\gamma = 0.577216...$ is Euler's number which results from the Born limit form of the ratio of the gamma functions. For zero quantum mass, we see that this gives an infinite phase correction due to the forward scattering singularity of the Coulomb problem. It is therefore convenient at this point to regularize the problem using a finite quantum mass model

$$\left(\frac{2}{1-\cos\vartheta}\right) \Rightarrow \frac{m_Q^2 + 4q^2}{m_Q^2 - M_X^2}$$

such that the phase is chosen to be given by

$$v(q) = \frac{1}{2} Z_1 Z_2 \alpha \frac{\mu}{q}\left[(1+4\gamma) - 2\log\left(\frac{m_Q^2 + 4q^2}{m_Q^2}\right)\right]$$

Therefore, a set of unitary models can be constructed which will give the Coulomb scattering behavior and reproduce the Rutherford scattering behavior in the Born limit for any functional form which satisfies

$$A(Z_1 Z_2 \alpha, q, \mu, m_Q) \underset{m_Q \to 0}{\Longrightarrow} Z_1 Z_2 \alpha \frac{\mu}{q}$$

5. 16

by using the form of equation 5. 12 and choosing the phase to be given by

$$\Delta(q,\vartheta) = \frac{1}{2} A(Z_1 Z_2 \alpha, q, \mu, m_Q) \left[ (1 + 4\gamma - 2\pi) + 2\log\left(\frac{m_Q^2}{m_Q^2 - M_X^2}\right) \right] + 2\tilde{\eta}(A(Z_1 Z_2 \alpha, q, \mu, m_Q))$$

5. 17

Any such model will exhibit all of the usual singular behaviors of Coulomb scattering in the zero quantum mass limit, including a forward scattering singularity and an essential singularity at zero momentum. However, any model of this form will satisfy the optical relationships for elastic scattering amplitudes given in Chapter III through equation 3. 7 and 3. 9. In the next section we construct a particular Lorentz invariant model which is used to illustrate the relationships of the previous chapter.

### C. A relativistic finite quantum mass model

#### 1. Single quantum type

We identify a momentum squared parameter in an analogous way with the previous example, but this time allow angular dependence in the transformation channel. We construct a model that consists of quantum exchange which generates the expected kinematic dependencies in the pp and pp-bar channels, but due to the summations over intermediate states, contributes many terms to the multipole expansion of the transformation (X) amplitude. The invariant momentum and energy parameters can be defined by

$$4Q^2(M_X, \xi_X) \equiv 2q_X^2(M_X^2)(1 + \zeta_{12} \xi_X) - M_X^2$$
$$\mathrm{M}^2(M_X, \xi_X) \equiv 4(m^2 + Q^2(M_X, \xi_X))$$

5. 18

Such an identification will result in many partial waves from the transformation channel which will coherently interfere with corresponding terms in the p-pbar channel. Therefore, we model a finite quantum mass unitary Coulombic amplitude using the form

$$A_X(M_X, \xi_X) = \frac{M(M_X, \xi_X)}{(2\pi m)^2} \frac{\sin \Delta(0, \xi_X)}{Q(M_X, \xi_X)} \left( \frac{m_Q^2 + 4Q^2(M_X, \xi_X)}{m_Q^2 - M_X^2} \right) e^{i\Delta(M_X^2, \xi_X)}$$

5. 19

where we can choose the phase factor $\Delta$ appropriately. Correspondence with Coulomb scattering places the following conditions on the parameters

$$A(Q(M_X^2, \xi_X), Z_1 Z_2 \alpha, m_Q) \underset{m_Q \to 0}{\Longrightarrow} Z_1 Z_2 \alpha \frac{m}{q}$$

$$\Delta(M_X^2, \xi_X) \underset{A \to 0}{\Longrightarrow} \frac{1}{2} A \left[ (1 + 4\gamma - 2\pi) + 2\log\left(\frac{m_Q^2}{m_Q^2 - M_X^2}\right) \right] + 2\tilde{\eta}(A)$$

5. 20

We want the partial waves to satisfy unitary relationships, which need not be guaranteed by forward scattering only, and we will assume that the phase factor has been chosen to make this true. As in the non-relativistic case, the form of this phase factor will have an off-shell extension that contains any bound state poles, and can be chosen to give a correspondence limit. Using the identification of equation 4. 2, either the particle-particle channel or the particle-antiparticle channel is immediately seen to be unitary, differing only in the relative signs of the two (distinguishable) charges.

$$A(M, \xi) = \frac{M}{(2\pi m)^2} \frac{\sin \Delta(0, \xi_X(M^2, \xi))}{q(M^2)} \left( \frac{m_Q^2 + 4q^2(M^2)}{m_Q^2 - M_X^2(M^2, \xi)} \right) e^{i\Delta(M_X^2(M^2, \xi), \xi_X(M^2, \xi))}$$

5. 21

The optical theorem can be directly checked in the form

$$\text{Im } A(M, \xi = 1) = -\pi \frac{m^2 q(M^2, m, m)}{M} \int_{-1}^{+1} |A(M, \xi)|^2 \, 2\pi \, d\xi$$

5. 22

with the particle-antiparticle channel satisfying the analogous form (with $Z_1 Z_2 < 0$)

$$\overline{A}(\overline{M},\overline{\xi}) = \frac{\overline{M}}{(2\pi m)^2} \frac{\sin\Delta(0,\xi_X(\overline{M}^2,\overline{\xi}))}{q(\overline{M}^2)} \left(\frac{m_Q^2 + 4q^2(\overline{M}^2)}{m_Q^2 - M_X^2(\overline{M}^2,\overline{\xi})}\right) e^{i\Delta(M_X^2(\overline{M}^2,\overline{\xi}),\xi_X(\overline{M}^2,\overline{\xi}))}$$

$$\operatorname{Im}\overline{A}(\overline{M},\overline{\xi}=1) = -\pi \frac{m^2 q(\overline{M}^2,m,m)}{\overline{M}} \int_{-1}^{+1} |\overline{A}(\overline{M},\overline{\xi})|^2 \, 2\pi d\overline{\xi}$$

5. 23

One thus observes that the constraint on one functional form written in terms of parameters in the transformation channel gives unitarity constraints in the two other channels. Particularly noteworthy is the observation that the identifications obtained give the expected form for the Coulomb scattering amplitudes for same charge and opposite charge scattering obtained using the nonrelativistic form, which gives a direct connection to actual physical interpretations of the derived amplitudes.

To obtain a unitary amplitude for identical particle-antiparticle elastic scattering and annihilation, one utilizes the results of the end of chapter 4 given by equations 4. 19-4. 21 to appropriately symmetrize and coherently add the p-pbar and X channel amplitudes to obtain the normalization constant and total amplitudes. The formulation insures that the result of these calculations will give the same results as an appropriately renormalized perturbative approach for the corresponding particles.

2. Two coupled quantum types

The previously demonstrated single quantum exchange model can be extended in numerous ways. We demonstrate an extension which will unitarily exchange two arbitrary mass quanta. This can be done using the 2-coupled channel parameterization of Stapp [11] by identifying the individual channels with the unitary scattering due to single quantum exchange. The overall scattering will then be unitary in a two-channel space, but the eigenvalues of that scattering matrix will themselves be individually unitary. Those eigenvalues satisfy

$$e^{2i\delta_\pm} = \frac{1}{2}\left[\cos 2w \left(e^{2i\delta_1} + e^{2i\delta_2}\right) \pm \sqrt{\cos^2 2w \left(e^{2i\delta_1} + e^{2i\delta_2}\right)^2 - 4e^{2i(\delta_1+\delta_2)}}\right]$$

5. 24

Therefore, given unitary amplitudes for quantum masses $m_{Q1}$ and $m_{Q2}$, this formula gives an overall amplitude which will be unitary for a model which has two quantum masses and corresponding coupling constants. The phase shifts are those corresponding to any unitary exchange of the $n^{th}$ quantum, where the amplitude for a given partial wave is given by

$$e^{2i\delta_n^J(\overline{M})} \equiv 1 + 2i\pi \rho_\pi(M, m_1, m_2) A_n^J(M)$$

5. 25

This defines two possible unitary amplitudes for a system which interacts via the exchange of 2 distinct quanta. The amplitudes are then given by the relationship

$$e^{2i\delta_\pm^J(\overline{M})} \equiv 1 + 2i\pi \rho_\pi(M, m_1, m_2) A_\pm^J(M)$$

5. 26

This type of formulation allows us to construct models with finite mass quanta which will be unitary, and thus not require renormalizability as a tenet of the model.

## VI. CONCLUSIONS

We have now accomplished our immediate goal of constructing a set of three Lorentz invariant amplitudes which preserve physical unitarity when used in the two-particle and particle-antiparticle sectors of a multi-body, multichannel formalism. In particular, we have found that the particle-particle, particle anti-particle, and transformation (e.g. $a\,\overline{a} \longleftrightarrow b\,\overline{b}$) channel amplitudes when written in terms of our invariant mass, angle parameters (i.e. $M, \xi$) satisfy *form invariance.* Consequently, if the appropriate unitarity constraint holds for one of them, the constraint which insures physical unitarity applies to the appropriate channels. Thus in model building, we can start from a model assumption for any one of the three amplitudes which we trust, and explore what the consequences of imposing particle-antiparticle symmetry (symbar) on that particular example will be. We have followed this though for a very simple model which may have little application, and for a class of models which are rich enough to give appropriate behavior for the scattering of charged particles and anti-particles, at least at low energy.

Clearly, our main goal of constructing a fully Lorentz invariant, unitary and cluster decomposable theory for a finite number of particles and quanta still remains to be accomplished. Out next paper will explore what happens when the type of amplitude constructed here is embedded in a three (or more) body space, and whether we can indeed describe quantum-particle scattering and particle-antiparticle pair production starting in that space.

**Acknowledgements**:

J.V.L. would like to acknowledge the hospitality of the Stanford Linear Accelerator Center during several periods of working out the details of these calculations. The authors wish to thank Walter Lamb for several careful readings and discussions of the manuscript. H.P.N. wishes to thank the director of SLAC for arranging the visitor program which enabled J.V.L. and H.P.N. to work out the final details of this paper.

1. M.Alfred, P.Kwizera, J.V.Lindesay, and H.P.Noyes, "A Non-Perturbative, Finite Particle Number Approach to Relativistic Scattering Theory", SLAC-PUB-8821, hep-th/0105241, (2001).
2. J.V.Lindesay, PhD thesis, available as SLAC Report No. SLAC-243 (1981).
3. J.V.Lindesay, A.J.Markevich, H.P.Noyes, and G.Pastrana, Phys.Rev.**D33**, 2339 (1986).
4. A.J.Markevich, "Relativistic Three-Particle Scattering Theory", PhD Thesis, Stanford University (1985).
5. A.J.Markevich, Phys.Rev.**D33**, 2350 (1986).
6. G.F.Chew, S Matrix Theory of Strong Interactions: a lecture note and reprint volume, Benjamin (1961).
7. G.F.Chew, The Analytic S Matrix: a basis for nuclear democracy, Benjamin (1966).
8. L.D.Faddeev, Zh.Eksp.Teor.Fiz.39,1459 (1960), Sov.Phys.-JETP **12**, 1014 (1961). See also L.D.Faddeev, Mathematical Aspects of the Three-Body Problem in Quantum Scattering Theory (Davey, New York, 1965). For the extension to the four-body problem, see e.g. O.A.Yakubovsky, Yad.Fiz. 5, (1967), Sov.J.Nucl.Phys. 5,937 (1967).
9. B.Lippman, Phys.Rev. **102**, 264 (1956).
10. M.L.Goldberger and K.M.Watson, Collision Theory, Wiley (1956), pp 197-209.
11. H.P.Stapp, T.J.Ypsilantis, and N.Metropolis, Phys.Rev. **105**, 302 (1957).
12. N.F.Mott and H.S.W.Massey,The Theory of Atomic Collisions, Oxford, New York (1949), 2nd edition, Ch III.
13. L.I.Schiff,Quantum Mechanics, McGraw-Hill (1955), Sec.20.